\documentclass[aps,twocolumn,showpacs,preprintnumbers,amsmath,amssymb,showkeys]{revtex4}

\usepackage{graphicx}
\usepackage{dcolumn}
\usepackage{bm}

\begin{document}

\preprint{CRPP-PSI/2004-UT-Theor05, 2-2013 PSI/SLS/RF}

\title{A Note on the Stability of Cable-in-Conduit Conductors with Current Dependent Power-Law Index}

\author{A. Anghel}

\affiliation{Paul Scherrer Institute, CH-5232 Villigen PSI, Switzerland}


\date{\today}

\begin{abstract}
A new stability criterion for cable-in-conduit superconductors with power-law current-voltage characteristic and current dependent power-law index, $n=n(I_c)$ is given. After a short discussion of the power-law volt-ampere characteristic, different models of stability are discussed with special stress on the differences to the older stability models. A typical stability case, the extended cryostability is analyzed in detail. This model is characterized by a smooth superconducting to normal transition and a power-law type heat generation. The change in the helium temperature, typical for a cable-in-conduit conductor, is included in the calculation. Finally, the connection and interrelation aspects between the two common approximations of the voltage-current characteristic of the technical superconductors, the power-law and the exponential function are investigated. It is shown that the exponential form is incompatible with the power-law form if the power-law index is a function of temperature and magnetic field. An alternative exponential form is proposed.
\end{abstract}

\pacs{23.23.+x,56.56.Dy}
\keywords{stability, superconductor, power-law, quench} 

\maketitle  

\section{Introduction}

Because the energy stored in a superconducting magnet, both
magnetic and mechanical, can easily be converted into heat,
upsetting the thermal equilibrium of the winding, the
superconductor temperature is difficult to be controlled and the
stability issue of superconductors became a crucial matter.
Therefore, the complete understanding of the thermal behavior of a
superconducting cable is of great importance both theoretically
and experimentally.

A crucial role in the stability issue is played by the way the
heat generation in the conductor takes place i.e. by the
volt-ampere characteristic (VAC) of the superconducting material.
With the advent of ceramic high temperature superconductors and
the ITER choice to use Cable-in-Conduit Conductors (CICC) for all
its magnets, the attention on the old non-linear VAC known as
Power-Law Conductor (PLC) has been refocused. It is expressed
analytically by the equation

\begin{equation}
\label{eq1}
E=E_c \left( {\frac{I}{I_c }} \right)^n
\end{equation}

where $E$ and $I$ are the electrical field and the operating
current and $E_c$ is a ``man-made'' ad-hoc voltage criterion used to
define the critical current $I_c $. The critical current is itself
a function of conductor temperature $T_{cond}$ and magnetic field
$B$. An important role is played here by the parameter $n$, appearing in the
exponent. It is responsible for the non-linearity of the volt-ampere characteristic and the standing heat generation in the real superconductors. With this model one can describe almost everything between a normal conductor ($n=1$) and a perfect superconductor ($n\to \infty$). If $n\to \infty $ the volt-ampere characteristic of ``classical Bean-model'' superconductors is recovered. For $n=1$ the classical resistor (Ohm's law) is obtained.

Based on experimental results we are
thought that $n$ is by no means always ``very high'' and it is by
no means constant. In high temperature superconducting (HTSC)
tapes it is maximum 15-20 but also values as low as 5 are
frequently reported in the literature. For low temperature
superconductors (LTS) $n$ can be very low (3-5) or medium high
(15-25) for Nb$_{3}$Sn conductors. Slightly higher values are
reported for NbTi based conductors. It is important to distinguish
between the strand $n$-value and the cable $n$-value.There are
only few cases when the two coincide. The cable $n$-values are
systematically lower that the strand values but there are also
exceptions to the rule. In some NbTi cables with low current
transfer capacity (insulated strands) the cable $n$-value was
higher that in the strands used to manufacture the cable
\cite{Stepanov:2003,Wesche:2003}.

The last issue concerning the index of the power law which is
probably known for a long time but was only recently recognized as
important, is the dependence of $n$ on temperature and magnetic
field. This dependence can be expressed in almost all cases as a
dependence on only one variable, the critical current $I_c =I_c
\left( {T_{cond} ,B} \right)$ and is characteristic for both
single strands and cables. For Nb$_{3}$Sn an additional dependence
on strain appears.

 The physical origin of this dependence is still obscure. Sure is that $n(I_c)$ is
 large at large $I_c $ (low temperature, low field or both) and low at low $I_c $
(higher temperature, field or both). Since both limits are well
defined thermally and magnetically it is not possible to assume
any other hidden factor which could explain this property.
Concerning the functional dependence of $n$ on $I_c $, linear, polynomial
and power-law dependencies have been reported or assumed. A
dependence of the form

\begin{equation}
\label{eq2}
n\left( {I_c } \right)=k I_c ^m
\end{equation}

with $m =0.3-0.5$ and a linear dependence

\begin{equation}
\label{eq3}
n\left( {I_c } \right)=a+bI_c
\end{equation}

will be considered here. They cover more or less much of the
spectrum of dependencies found in the literature.

The paper is organized as follows. In II we first review different
one-dimensional (1D) stability models and investigate the impact
of changing from the composite parallel circuit model of heat
generation to the heat generation by the power-law (also known as
index heating). In III a stability model of the unconditional type
with a PLC-type VAC is investigated. Finally, in IV the relation
between the PLC with variable n-index and the exponential form of
VAC is investigated.

\section{Models of Stability}

With the reference to a superconducting strand, the following constituents
are relevant for stability. The conductor itself, composed of superconducting
filaments and the stabilizing matrix (copper), and the coolant (liquid or
supercritical He for LTS, LN2 for HTSC).

If we adopt for simplicity a 1D model, the conductor $T_{cond}$ and helium $T_{he}$ temperatures are governed by the following equations

\begin{equation}
\label{eq4} 
\begin{split}
\rho_{cond} C_{cond} A_{cond} \frac{\partial
T_{cond}}{\partial t}=\frac{\partial }{\partial x}\left( {\kappa
A_{cond} \frac{\partial T_{cond} }{\partial x}} \right) \\ 
+G\left({T_{cond} ,t} \right) 
-hp_w \left( {T_{cond} -T_{he} }\right)+P_{pulse}\left( t \right)  \\
\rho_{he} C_{he} A_{he} \frac{\partial T_{he} }{\partial t}+\dot
{m}C_{he} \frac{\partial T_{he} }{\partial x}=hp_w
\left({T_{cond} -T_{he} } \right)\qquad
\end{split}
\end{equation}

 where $G\left( {T_{cond} ,t} \right)=E\left( {T_{cond} ,t}
\right)I\left( t \right)$ is the heat generated in the
superconductor and $P_{pulse} \left( t \right)$ is the pulse
creating a deviation from equilibrium. The transient is initiated
either by the time dependence of the electric field, of the current or from an
external heat pulse. Depending on what terms are retained in the above equations and how the electrical field is represented as a function of current, different models of stability can be defined. This is illustrated in Table 1 and 2 for the conductor and helium temperature equations. The terms taken into account in each case are marked by the $\times$ sign and the resulting model name is given in the last column. Let us start with the case when we completely neglect the helium equation. We assume that the helium temperature is constant $T_{he} =T_b $ and adopt the parallel circuit model for the heat generation in the composite \cite{Iwasa:1994} i.e. the heat generation is formulated as

\squeezetable
\begin{table*}[htb]
\caption{\label{tab:table1} Models for stability, conductor
equations.}
\begin{center}
\begin{ruledtabular}
\begin{tabular} {ccccc}                
Heat Accumulation& Heat Conduction& Internal Heat Generation& Heat
\par Exchange& Model \\
 \hline \\
$\rho_{cond}C_{cond}A_{cond} \dfrac{\partial T_{cond}}{\partial t}$& $\dfrac{\partial }{\partial x}\left( {\kappa
A_{cond} \dfrac{\partial T_{cond} }{\partial x}} \right)$& $EI$&
$hp_w \left( {T_{cond} -T_{he} } \right)$& \\
\\
\hline
&
&
$\times $&
$\times $&
Cryostability \\
&
$\times $&
$\times $&
$\times $&
Equal area \\
&
$\times $&
$\times $&
&
MPZ \\
$\times $&
$\times $&
$\times $&
$\times $&
Dynamic stability \\
$\times $& &
$\times $& &
Protection \\
$\times $&
$\times $&
$\times $&
&
Adiabatic NZP \\
\end{tabular}
\end{ruledtabular}
\end{center}
\end{table*}

\squeezetable
\begin{table*}[tb]
\caption{\label{tab:table2} Models of stability, helium equations.}
\begin{center}
\begin{ruledtabular}
\begin{tabular}  {ccccc}                       
Heat Accumulation& Convection& Heat Exchange& Constant Temperature& Model \\
\hline \\ 
$\rho_{he}C_{he}A_{he} \dfrac{\partial T_{he} }{\partial t}$ & $\dot {m}C_{he} \dfrac{\partial T_{he} }{\partial x}$ & $hp_w\left( {T_{cond} -T_{he} } \right)$ & $T_{he} =T_b $& \\
\\
\hline
& & &$\times $&Pool cooling \\
$\times $& &$\times $& &Generalized transient \\
$\times $&
$\times $&
$\times $&
&
CICC \\
\end{tabular}
\end{ruledtabular}
\end{center}
\end{table*}

\begin{widetext}
\begin{equation}
\label{eq5} 
G^\infty \left( {T_{cond} ,t} \right)=\left\{
{{\begin{array}{*{25}c}
 0 \hfill & {\textrm{for}} \hfill & {T_{cond} <T_{cs} } \hfill \\ \\
 {\dfrac{\rho _{cu} }{A_{cu} }I^2\left( {\dfrac{T_{cond} -T_{cs} }{T_c -T_{cs}}} \right)} \hfill & {\textrm{for}} \hfill & {T_{cs} <T_{cond}<T_c }\hfill \\ \\
 {\dfrac{\rho _{cu} }{A_{cu} }I^2} \hfill & {\textrm{for}} \hfill & {T_{cond} >T_c } \hfill \\
\end{array} }} \right.
\end{equation}
\end{widetext}

In this case we recover the whole class of old models of stability for bath
cooled magnets and "perfect superconductors" with $n=\infty $.

For instance, if we assume stationary conditions by setting the time
derivatives in Table 1 and Table 2 to zero we recover the classical
"cryostability" or "Stekly" model \cite{Kantrowitz:1965,Stekly:1969}

\begin{equation}
\label{eq6}
G^\infty \left( {T_{cond} } \right)=hp_w \left( {T_{cond} -T_b } \right)
\end{equation}

the "equal area" model of Maddock, James and Norris \cite{Maddock:1969}

\begin{equation}
\label{eq7} 
\frac{\partial }{\partial x}\left( {\kappa A_{cond}
\frac{\partial T_{cond} }{\partial x}} \right)+G^\infty \left(
{T_{cond} } \right)=hp_w \left( {T_{cond} -T_b } \right)
\end{equation}

or the "minimum propagating zone" (MPZ) model, Martinelli and Wipf
\cite{Martinelli:1997}

\begin{equation}
\label{eq8} 
\frac{\partial }{\partial x}\left( {\kappa
A_{cond}\frac{\partial T_{cond} }{\partial x}} \right)+G^\infty
\left( {T_{cond} } \right)=0
\end{equation}

by appropriately choosing two of the three terms left in
Eq.(\ref{eq4}). 

Most of the properties of the above models are due
to a particularity of the functional form of the heat generation term as expressed by
Eq.(\ref{eq5}). As can be seen, the heat generation in the strand is exactly zero
below a certain temperature, $T_{cs} $ -the current-sharing
temperature. With the finite $n$ conductors, where $G$ is non zero at all
temperatures, this property does not hold and drastically changes
are expected. A characteristic of power-law conductor models is
that there is always power generation at all temperatures.

The cryostability model Eq.(\ref{eq6}), has an additional
particularity. Due to the disappearance of the conduction term, it
is not a differential equation anymore and only local values are
involved. It can be derived directly from a 0D model by
neglecting the time derivative.

If the time dependence is not neglected, Eq.(\ref{eq4}) gives
other two models for stability. First we have the dynamic
stability model of Hart \cite{Hart:1969}

\begin{equation}
\label{eq9}
\begin{split}
 \rho_{cond}C_{cond}A_{cond} \frac{\partial T_{cond}
}{\partial t}=\frac{\partial }{\partial x}\left( {\kappa A_{cond}
\frac{\partial T_{cond} }{\partial x}} \right) \\ +G^\infty \left(
{T_{cond} ,t} \right)-hp_w \left( {T_{cond} -T_{he} } \right)
\end{split}
\end{equation}

where practically all terms are considered, then the so called
"protection" model \cite{Maddock:1968}

\begin{equation}
\label{eq10} 
\rho_{cond}C_{cond}A_{cond} \frac{\partial T_{cond}
}{\partial t}=G^\infty \left( {T_{cond} ,t} \right)
\end{equation}

which completely neglects cooling and conduction effects. Finally
we have also the adiabatic normal zone propagation (NZP) model
\cite{Rakhmanov:2000}

\begin{equation}
\label{eq11}
\begin{split}
\rho_{cond}C_{cond}A_{cond} \frac{\partial T_{cond}
}{\partial t}=\frac{\partial }{\partial x}\left( {\kappa A_{cond}
\frac{\partial T_{cond} }{\partial x}} \right) \\ +G^\infty \left(
{T_{cond} ,t} \right)
\end{split}
\end{equation}

where only the cooling effect is neglected.

For CICC there is a finite amount of helium at each cross section
of the cable and the second part of Eq.(\ref{eq4}) cannot be
neglected. Also heat generation in CICC is better described by
power-law. With these too facts in mind we can start building
stability models for CICC picking terms in Table 1 and 2. Again, first assuming  stationary
conditions, we have the following stationary models:

a) \underline {Extended Cryostability Model}

\begin{eqnarray}
\label{eq12}
 G\left( {T_{cond} ,t} \right)=hp_w \left( {T_{cond} -T_{he} } \right) \nonumber \\
 \dot {m}C_{he} \frac{\partial T_{he} }{\partial x}=hp_w \left( {T_{cond}
-T_{he} } \right)
 \end{eqnarray}

b) \underline {Extended Equal Area Model}

\begin{eqnarray}
\label{eq13}
 \frac{\partial }{\partial x}\left( {\kappa A_{cond} \frac{\partial T_{cond}}{\partial x}} \right)+G\left(T_{cond},t\right)=\nonumber\\=hp_w \left(T_{cond}-T_{he}\right) \nonumber \\
 \dot{m}C_{he} \frac{\partial T_{he} }{\partial x}=hp_w \left(T_{cond}-T_{he}\right)
 \end{eqnarray}

In the class of time dependent models we can imagine the following
case

c) \underline {0D Transient Stability Model }

\begin{eqnarray}
\label{eq14} \rho_{cond}C_{cond}A_{cond} \frac{\partial T_{cond}
}{\partial t} = G\left( {T_{cond} ,t} \right) \nonumber \\
-hp_w \left( {T_{cond}-T_{he} } \right)+P_{pulse} \left( t \right) \nonumber \\
 \rho_{he}C_{he}A_{he} \frac{\partial T_{he} }{\partial t} = hp_w \left( {T_{cond} -T_{he} }
\right)
 \end{eqnarray}

but as discussed above it make sense only for $ G\equiv G^\infty $.

\section{Smooth Superconducting to Normal Transition}

In this section we will concentrate mainly on the extended
cryostability model described by Eq.(\ref{eq12}) and assume a
power-law conductor with a voltage-current characteristic
described by Eq.(\ref{eq1}). In other words we consider the helium
temperature as variable and assume a smooth transition from the
superconducting to the normal state. Additionally and opposite to
other models \cite{Rakhmanov:2000,Anghel:2003} we will consider that the index $n$ is a
function of the critical current and not a constant. The model
equations with the explicit temperature dependence of $I_c$ are

\begin{eqnarray}
\label{eq15}
 E_c I\left( {\frac{I}{I_c \left( {T_{cond} } \right)}} \right)^{n\left(
{I_c } \right)}=hp_w \left( {T_{cond} -T_{he} } \right) \nonumber \\
 \dot {m}C_{he} \frac{\partial T_{he} }{\partial x}=hp_w \left( {T_{cond}
-T_{he} } \right)
 \end{eqnarray}

a system of two equation, one partial-differential and the other one pure
algebraic. This system can be reduced to a single partial-differential
equation as follows. We use the first equation in
(\ref{eq15}) to express the helium temperature as a function of
conductor temperature

\begin{equation}
\label{eq16}
T_{he} =T_{cond} -\frac{E_c I}{hp_w }\left( {\frac{I}{I_c \left( {T_{cond} }
\right)}} \right)^{n\left( {I_c } \right)}
\end{equation}

and then we substitute it in the second equation. Calculating the
derivative $ \partial T_{he}/\partial x$ is the only difficult part.

\begin{eqnarray}
\label{eq17}
 \frac{\partial T_{he} }{\partial x}=\left(\frac{\partial T_{he} }{\partial
T_{cond} }\right)\left(\frac{\partial T_{cond} }{\partial x}\right)= \nonumber \\
=\left[ {1-\frac{E_c I}{hp_w}\frac{\partial }{\partial T_{cond} }\left( {\frac{I}{I_c \left( {T_{cond} }
\right)}} \right)^{n\left( {I_c } \right)}} \right]\frac{\partial T_{cond}}{\partial x}= \nonumber \\
 =\left[ {1-\frac{G}{hp_w }\left( {\frac{\partial n}{\partial I_c }\ln
\frac{I}{I_c }-\frac{n}{I_c }} \right)\frac{\partial I_c }{\partial T_{cond}
}} \right]\frac{\partial T_{cond} }{\partial x} \nonumber \\
 \end{eqnarray}

Finally substituting Eq.(\ref{eq17}) in Eq.(\ref{eq15}) and using the first equation again
we get finally

\begin{eqnarray}
\label{eq18} 
\dot{m} C_p \frac{\partial T_{cond} }{\partial x}=G\times \nonumber \\
\times \left[ {1-\frac{G}{hp_w }\left( {\frac{\partial n}{\partial I_c
}\ln \left( {\frac{I}{I_c }} \right)-\frac{n}{I_c }}
\right)\frac{\partial I_c }{\partial T_{cond} }} \right]^{-1}
\end{eqnarray}

an equation for the conductor temperature $T_{cond} $ alone. This can be integrated numerically and the conductor temperature profile can be determined. This is however not our goal here. We are interested here much more in finding out the conditions under which a solution of Eq.(\ref{eq18}) does exists.

First let us observe that ${\partial T_{cond} } \mathord{\left/
{\vphantom {{\partial T_{cond} } {\partial x}}} \right.
\kern-\nulldelimiterspace} {\partial x}$ cannot be zero since for
a power-law conductor $G$ is never zero. Secondly, it is clear
physically that ${\partial T_{cond} } \mathord{\left/ {\vphantom
{{\partial T_{cond} } {\partial x}}} \right.
\kern-\nulldelimiterspace} {\partial x}$ should be positive. The
conductor temperature increases from some lower value at the inlet
of the modeled section and reaches a maximum at its end due to the
heat accumulated in the helium which is then convectively
transported away along the conductor. Therefore a nontrivial
solution of Eq.(\ref{eq18}) is possible only if the following
condition is satisfied

\begin{equation}
\label{eq19}
1-\frac{G}{hp_w }\left[ {\frac{\partial n}{\partial I_c }\ln \left(
{\frac{I}{I_c }} \right)-\frac{n}{I_c }} \right]\frac{\partial I_c
}{\partial T_{cond} }\geqslant 0
\end{equation}

From Eq.(\ref{eq19}), expressing $G$ as $EI_{q}$ where $I_{q}$ is
the quench current, we arrive at the stability condition 

\begin{equation}
\label{eq20} E\leqslant E_q =hp_w \left[ I_q \left(
{\frac{\partial n}{\partial I_c }\ln \left( {\frac{I_q }{I_c }}
\right)-\frac{n}{I_c }} \right)\frac{\partial I_c }{\partial
T_{cond} } \right]^{-1}
\end{equation}

i.e the conductor with current dependent power-law index is stable against thermal perturbations if the local electrical field is smaller than $E_q$ the take-off or quench electrical field.

From experimental data we know that ${\alpha =\partial n}
\mathord{\left/ {\vphantom {{\alpha =\partial n} {\partial I_c
>0}}} \right. \kern-\nulldelimiterspace} {\partial I_c >0}$ and
${\beta =\partial I_c } \mathord{\left/ {\vphantom {{\beta
=\partial I_c } {\partial T_{cond} <0}}} \right.
\kern-\nulldelimiterspace} {\partial T_{cond} <0}$. The parameters
$\alpha $ and $\beta $ are slowly varying functions of temperature
and field and can be considered in the first order of
approximation as constants. This is true in particular for the
linear dependence of $n$ on $I_c $ which is a good approximation
for high temperature and/or fields i.e. for small critical
currents. For constant $n$ , Eq.(\ref{eq20}) simplifies to

\begin{equation}
\label{eq21} E_{q}=\frac{hp_wI_c}{I_q n} \left| \frac{\partial
I_c}{\partial T_{cond}} \right|^{-1}=\frac{hp_wI_c}{I_qn\beta}
\end{equation}

Expressing the power-law relation, Eq.(\ref{eq1}, at the quench point: $E=E_q$ at $I=I_q$

\begin{equation}
\label{eq22} 
E_q=E_c \left(\frac{I_q}{I_c}\right)^n
\end{equation}

 one arrives finally at

\begin{equation}
\label{eq23} E_q=E_c
\left(\frac{hp_w}{nE_c\beta}\right)^{\frac{n}{n+1}}\mathop  \to
\limits_{n \to \infty }\frac{hp_w}{n\beta}
\end{equation}

a relation obtained previously in \cite{Rakhmanov:2000,Anghel:2003}. Equation (\ref{eq20}) is therefore the generalized expression for the quench condition in a conductor with variable power-law index $n$ 

\section{Relations to the Exponential Form of the Volt-Ampere Characteristic}
An alternative form of the volt-ampere characteristic, used mainly
in the Russian literature on this subject, is given by the following
exponential form \cite{Dorofejev:1980}, \footnote{In this paper the VAC is given in the equivalent form, $E=IR_n\exp \left( -T_c/T_0+T/T_0+B/B_0+I/I_0\right)$ where $R_n$ is the superconductor normal resistance and $T_c$ is the critical temperature.}.

\begin{eqnarray}
\label{eq24}
E\left(T_{cond},B,I\right)=\nonumber \\ 
E_c\exp\left[\frac{I-I_c\left(T_b,B_b\right)}{I_0}  
+\frac{T_{cond}-T_b}{T_0}+\frac{B-B_b}{B_0}\right]
\end{eqnarray}

where $T_b$ and $B_b $ are a reference temperature and a
reference magnetic field. The reference point is chosen such that
$E=E_c$ at $T_{cond}=T_b$, $B=B_b$ and $I=I_c \left( {T_b ,B_b }
\right)$. The constants $I_0$ ,$T_0$ and $B_0$ are growth 
parameters of $E$ by the current, temperature and magnetic field \footnote{ in \cite{Dorofejev:1980} $I_0$, $T_0$ and $B_0$ are named "increasing" parameters for current, temperature and field.}. It is
implicit assumed, but never explicitly stated, that these
parameters are really constant i.e. they do not depend on
temperature, current and field. The whole dependence on temperature,
current and field is exactly what is seen in Eq.(\ref{eq24}) explicitly and
nothing else more.

It the following we will establish the relation between this
form and the power-law form of Eq.(\ref{eq1}) which for convenience is written here
once again

\begin{equation}
\label{eq25} E\left( {T_{cond} ,B,I} \right)=E_c \left(
{\frac{I}{I_c \left( {T_{cond} ,B} \right)}} \right)^{n\left( {I_c
} \right)}
\end{equation}

First let us observe a fundamental difference between the two
functional forms. In the exponential form, the temperature and
magnetic field dependence is explicit while in the power-law form
it is implicit, through the temperature and field dependence of
$I_c$. The connection between the two forms can be found by taking
the first derivative of Eq.(\ref{eq24}) with respect to $T_{cond}$
at constant current and magnetic field. The result is

\begin{equation}
\label{eq26} 
\left(\frac{\partial E}{\partial T_{cond} }\right)_{I,B}=\frac{E}{T_0 }
\end{equation}

showing that indeed $1 \mathord{\left/ {\vphantom {1 {T_0 }}}
\right. \kern-\nulldelimiterspace} {T_0 }$ is the growth factor of
an exponential grow. The same derivative calculated from
Eq.(\ref{eq25}) gives

\begin{equation}
\label{eq27} \frac{\partial E}{\partial T_{cond} }=E\left[
{\frac{\partial n}{\partial I_c }\ln \left( {\frac{I}{I_c }}
\right)-\frac{n}{I_c }} \right]\frac{\partial I_c }{\partial
T_{cond} }
\end{equation}

Now, as it is well known, the numerical difference between these two
functional forms is vanishing small within $2-3$ orders of
magnitude and therefore we can equate the results from
Eq.(\ref{eq26}) and Eq.(\ref{eq27}) with the final result

\begin{equation}
\label{eq28} \frac{1}{T_0 }=\left[ {\frac{\partial n}{\partial I_c
}\ln \left( {\frac{I}{I_c }} \right)-\frac{n}{I_c }}
\right]\frac{\partial I_c }{\partial T_{cond} }
\end{equation}

showing that the growth parameter is related to the slope of the
critical current as a function of temperature and of $n$-index as
a function of critical current. From Eq.(\ref{eq28}) it is easy to
see that a variable $n$ makes $T_0 $ an explicit function of
current and an implicit function of temperature and field through
the dependence on $I_c$. This contradicts the assumption on which
expansion in Eq.(\ref{eq24}) is based i.e. that $T_0$ should be a
constant and not a function of anything. Therefore the exponential
functional form is not compatible to the power-law with a variable
index. For constant $n$, i.e. for $ \partial n/\partial
I_c=0 $ and at a fixed temperature and field such that also
$\partial I_c/\partial T_{cond}<0$ is constant, we get an
expression which is independent of current

\begin{equation}
\label{eq29} \frac{1}{T_0 }=-\frac{n}{I_c }\left(\frac{\partial
I_c }{\partial T_{cond} }\right)=\frac{n}{I_c }\left|
{\frac{\partial I_c }{\partial T_{cond} }} \right|
\end{equation}

a first relation between the parameters of the two functions. As
can be seen the temperature growth parameter is connected to the
$n$-index and the logarithmic derivative of the critical current.
A similar relation is obtained immediately for $I=I_c$ from Eq.(\ref{eq28}) without any
other supplementary conditions.

For the magnetic field dependence we get using the same procedure,
i.e. equating the derivatives ${\partial E} \mathord{\left/
{\vphantom {{\partial E} {\partial B}}} \right.
\kern-\nulldelimiterspace} {\partial B}$ of Eq.(\ref{eq24}) and
Eq.(\ref{eq25})

\begin{equation}
\label{eq30} \frac{1}{B_0 }=\left[ {\frac{\partial n}{\partial I_c
}\ln \left( {\frac{I}{I_c }} \right)-\frac{n}{I_c }}
\right]\frac{\partial I_c }{\partial B}
\end{equation}

which is again current dependent through the logarithmic term. For
constant $n$ or at $I=I_c $ we get

\begin{equation}
\label{eq31} \frac{1}{B_0 }=\frac{n}{I_c }\left| {\frac{\partial
I_c }{\partial B}} \right|
\end{equation}

Finally, for the current growth factor the following relation is
obtained

\begin{equation}
\label{eq32} \frac{1}{I_0 }=\frac{n}{I}
\end{equation}

which at $I=I_c $

\begin{equation}
\label{eq33} \frac{1}{I_0 }=\frac{n}{I_c }
\end{equation}

It can be seen that in principle all growth parameter violate the
assumption of complete separability of temperature, current and
field dependence if the index of the power law depends on
temperature and field. Therefore, in general, the exponential form
is incompatible with a power-law with variable index. However, at
$I=I_c $ and/or at fixed temperature and field the two expressions
are in agreement.

The expressions for the growth parameters at $I=I_c $ can be used
to re-express the exponential form. Substituting Eqs.(\ref{eq29})
,(\ref{eq31}) and (\ref{eq33}) in Eq.(\ref{eq24}) we obtain

\begin{widetext}
\begin{eqnarray}
\label{eq.34}
 E=E_c \exp \left[ {\left( {I-I_{b} } \right)\frac{n}{I_c }-\left( {T-T_b
} \right)\frac{n}{I_c }\frac{\partial I_c }{\partial T}-\left( {B-B_b }
\right)\frac{n}{I_c }\frac{\partial I_c }{\partial B}} \right]= \nonumber \\
 =E_c \exp \left[ {\frac{n}{I_c }\left( {I-I_{b} -\left( {T-T_b }
\right)\frac{\partial I_c }{\partial T}-\left( {B-B_b }
\right)\frac{\partial I_c }{\partial B}} \right)} \right]
 \end{eqnarray}
\end{widetext}

where $I_{b} =I_c \left( {T_b ,B_b } \right)$ and $I_c =I_c
\left( {T,B} \right)$. Observing now that the last three terms in
the exponential represent the Taylor series development of $I_c
\left( {T,B} \right)$ around the point $\left( {T_b ,B_b }
\right)$ we get an expression for the exponential form where the
power-law index appears explicitly

\begin{equation}
\label{eq35} E=E_c \exp \left[ {\frac{n}{I_c }\left( {I-I_c }
\right)} \right]
\end{equation}

which can be used as an alternative to the standard exponential
form Eq.(\ref{eq24}) if the power-law index depends on temperature
and field. Further, it can be shown that the new expression is
equivalent to a power-law function to order ${O(1} \mathord{\left/
{\vphantom {{O(1} {n^2}}} \right. \kern-\nulldelimiterspace}
{n^2})$. For this purpose we first invert Eq.(\ref{eq35})

\begin{equation}
\label{eq36} I=I_c \left( {1+\frac{1}{n}\ln \left( {\frac{E}{E_c
}} \right)} \right)
\end{equation}

The inverted form of the power-law function is

\begin{equation}
\label{eq37} I=I_c \left( {\frac{E}{E_c }}
\right)^{\frac{1}{n}}\approx I_c \left( {1+\frac{1}{n}\ln \left(
{\frac{E}{E_c }} \right)+O\left( {\frac{1}{n^2}} \right)\ldots }
\right)
\end{equation}

where on the right-hand side we have developed the function in
powers of $1 \mathord{\left/ {\vphantom {1 n}} \right.
\kern-\nulldelimiterspace} n$ up to the second order. The
logarithmic function is a slowly varying function and for large
$n$, we see immediately that to order $1 \mathord{\left/
{\vphantom {1 {n^2}}} \right. \kern-\nulldelimiterspace} {n^2}$,
Eq.(\ref{eq36}) and Eq.(\ref{eq37}) are identical.

In conclusion, the power-law form and the exponential form are
identical to order ${O(1} \mathord{\left/ {\vphantom {{O(1}
{n^2}}} \right. \kern-\nulldelimiterspace} {n^2})$. For
applications where the temperature and field dependence of the
$n$-index is essential, the form in Eq.(\ref{eq35}) is
recommended, whereas for currents close to critical current, the
standard parametric form, Eq.(\ref{eq25}), can be used.

The quench condition expressed with the help of the exponential
form can be easily obtained from Eq.(\ref{eq21}) and
Eq.(\ref{eq29})

\begin{equation}
\label{eq38} E_q=\frac{hp_wT_0}{I_q}
\end{equation}

a very simple and elegant relation between the local electric
field at quench, the quench current and the temperature growth
parameter. However, the applicability of this equation is limited
to cases with constant or slowly varying $n$ as discussed before.

\bibliography{stability}

\begin{thebibliography}{12}
\expandafter\ifx\csname natexlab\endcsname\relax\def\natexlab#1{#1}\fi
\expandafter\ifx\csname bibnamefont\endcsname\relax
  \def\bibnamefont#1{#1}\fi
\expandafter\ifx\csname bibfnamefont\endcsname\relax
  \def\bibfnamefont#1{#1}\fi
\expandafter\ifx\csname citenamefont\endcsname\relax
  \def\citenamefont#1{#1}\fi
\expandafter\ifx\csname url\endcsname\relax
  \def\url#1{\texttt{#1}}\fi
\expandafter\ifx\csname urlprefix\endcsname\relax\def\urlprefix{URL }\fi
\providecommand{\bibinfo}[2]{#2}
\providecommand{\eprint}[2][]{\url{#2}}

\bibitem[{\citenamefont{B.Stepanov et~al.}(2003)\citenamefont{B.Stepanov,
  A.Anghel, P.Bruzzone, and M.Vogel}}]{Stepanov:2003}
\bibinfo{author}{\bibnamefont{B.Stepanov}},
  \bibinfo{author}{\bibnamefont{A.Anghel}},
  \bibinfo{author}{\bibnamefont{P.Bruzzone}}, \bibnamefont{and}
  \bibinfo{author}{\bibnamefont{M.Vogel}}, \bibinfo{journal}{MT-18, Morioka,
  Japan}  (\bibinfo{year}{2003}).

\bibitem[{\citenamefont{R.Wesche et~al.}(October 2003)\citenamefont{R.Wesche,
  A.Anghel, B.Stepanov, and P.Bruzzone}}]{Wesche:2003}
\bibinfo{author}{\bibnamefont{R.Wesche}},
  \bibinfo{author}{\bibnamefont{A.Anghel}},
  \bibinfo{author}{\bibnamefont{B.Stepanov}}, \bibnamefont{and}
  \bibinfo{author}{\bibnamefont{P.Bruzzone}}, \bibinfo{journal}{MT-18, Morioka,
  Japan}  (\bibinfo{year}{October 2003}).

\bibitem[{\citenamefont{Iwasa}(1994)}]{Iwasa:1994}
\bibinfo{author}{\bibfnamefont{Y.}~\bibnamefont{Iwasa}},
  \emph{\bibinfo{title}{Case Studies in Superconducting Magnets}}
  (\bibinfo{publisher}{Plenum Press, New York and London},
  \bibinfo{year}{1994}).

\bibitem[{\citenamefont{A.R.Kantrowitz and
  Z.J.J.Stekly}(1965)}]{Kantrowitz:1965}
\bibinfo{author}{\bibnamefont{A.R.Kantrowitz}} \bibnamefont{and}
  \bibinfo{author}{\bibnamefont{Z.J.J.Stekly}}, \bibinfo{journal}{Appl. Phys.
  Lett.} \textbf{\bibinfo{volume}{6}}, \bibinfo{pages}{56}
  (\bibinfo{year}{1965}).

\bibitem[{\citenamefont{Z.J.J.Stekly et~al.}(1969)\citenamefont{Z.J.J.Stekly,
  R.Thome, and B.Trauss}}]{Stekly:1969}
\bibinfo{author}{\bibnamefont{Z.J.J.Stekly}},
  \bibinfo{author}{\bibnamefont{R.Thome}}, \bibnamefont{and}
  \bibinfo{author}{\bibnamefont{B.Trauss}}, \bibinfo{journal}{J.Appl.Phys.}
  \textbf{\bibinfo{volume}{40}}, \bibinfo{pages}{2238} (\bibinfo{year}{1969}).

\bibitem[{\citenamefont{B.J.Maddock et~al.}(1969)\citenamefont{B.J.Maddock,
  G.B.James, and W.T.Norris}}]{Maddock:1969}
\bibinfo{author}{\bibnamefont{B.J.Maddock}},
  \bibinfo{author}{\bibnamefont{G.B.James}}, \bibnamefont{and}
  \bibinfo{author}{\bibnamefont{W.T.Norris}}, \bibinfo{journal}{Cryogenics}
  \textbf{\bibinfo{volume}{9}}, \bibinfo{pages}{261} (\bibinfo{year}{1969}).

\bibitem[{\citenamefont{A.P.Martinelli and S.L.Wipf}(1997)}]{Martinelli:1997}
\bibinfo{author}{\bibnamefont{A.P.Martinelli}} \bibnamefont{and}
  \bibinfo{author}{\bibnamefont{S.L.Wipf}}, \bibinfo{journal}{Proc. Appl.
  Supercond. Conf., IEEE Pub.} \textbf{\bibinfo{volume}{72CHO682-5-TABSC}},
  \bibinfo{pages}{3311} (\bibinfo{year}{1997}).

\bibitem[{\citenamefont{Hart}(1969)}]{Hart:1969}
\bibinfo{author}{\bibfnamefont{H.}~\bibnamefont{Hart}}, \bibinfo{journal}{Proc.
  1968 Summer Study on Superconductive Devices and Accelerators} p.
  \bibinfo{pages}{571} (\bibinfo{year}{1969}).

\bibitem[{\citenamefont{Maddock and James}(1968)}]{Maddock:1968}
\bibinfo{author}{\bibfnamefont{B.~J.} \bibnamefont{Maddock}} \bibnamefont{and}
  \bibinfo{author}{\bibfnamefont{G.~B.} \bibnamefont{James}},
  \bibinfo{journal}{Proc. Inst. Electr. Eng.} \textbf{\bibinfo{volume}{115}},
  \bibinfo{pages}{543} (\bibinfo{year}{1968}).

\bibitem[{\citenamefont{A.L.Rakhmanov et~al.}(2000)\citenamefont{A.L.Rakhmanov,
  V.S.Vysotsky, Yu.A.Ilyn, T.Kiss, and M.Takeo}}]{Rakhmanov:2000}
\bibinfo{author}{\bibnamefont{A.L.Rakhmanov}},
  \bibinfo{author}{\bibnamefont{V.S.Vysotsky}},
  \bibinfo{author}{\bibnamefont{Yu.A.Ilyn}},
  \bibinfo{author}{\bibnamefont{T.Kiss}}, \bibnamefont{and}
  \bibinfo{author}{\bibnamefont{M.Takeo}}, \bibinfo{journal}{Cryogenics}
  \textbf{\bibinfo{volume}{40}}, \bibinfo{pages}{19} (\bibinfo{year}{2000}).

\bibitem[{\citenamefont{A.Anghel}(2003)}]{Anghel:2003}
\bibinfo{author}{\bibnamefont{A.Anghel}}, \bibinfo{journal}{Cyrogenics}
  \textbf{\bibinfo{volume}{43}}, \bibinfo{pages}{225} (\bibinfo{year}{2003}).

\bibitem[{\citenamefont{G.L.Dorofejev et~al.}(1980)\citenamefont{G.L.Dorofejev,
  A.B.Imenitov, and E.Yu.KLimenko}}]{Dorofejev:1980}
\bibinfo{author}{\bibnamefont{G.L.Dorofejev}},
  \bibinfo{author}{\bibnamefont{A.B.Imenitov}}, \bibnamefont{and}
  \bibinfo{author}{\bibnamefont{E.Yu.KLimenko}}, \bibinfo{journal}{Cryogenics}
  \textbf{\bibinfo{volume}{20}}, \bibinfo{pages}{307} (\bibinfo{year}{1980}).

\end{thebibliography}

\end{document}